\documentstyle[12pt,epsfig]{article}
\newcommand{\beq}{\begin{equation}}
\newcommand{\eeq}{\end{equation}}
\newcommand{\bea}{\begin{eqnarray}}
\newcommand{\eea}{\end{eqnarray}}

\begin{document}

\begin{titlepage}
\begin{flushright}
cond-mat/0209146
\end{flushright}
\vskip2.8cm
\begin{center}
{\LARGE
Bayesian Entropic Inverse Theory Approach     
\vskip0.5cm
to Implied Option Pricing with Noisy Data
}
\vskip1.5cm
{\Large Igor Halperin} 
\vskip0.5cm
        BFM Financial Research \\
\vskip1.0cm
\today
\vskip1.0cm
{\small e-mail: $i\_halperin@hotmail.com $}
\vskip1.0cm
{\Large Abstract:\\}
\end{center}
\parbox[t]{\textwidth}{
A popular approach to nonparametric option pricing is the 
Minimum Cross Entropy (MCE) method based on minimization of 
the relative Kullback-Leibler entropy of the price density distribution
and a given reference density, with observable option
prices serving as constraints.  When market prices are noisy,
the MCE method tends to overfit the data and often becomes unstable.  
We propose a non-parametric option pricing method whose input are 
noisy market prices of arbitrary number of European options with 
arbitrary maturities. Implied transition densities are calculated using 
the Bayesian inverse theory with entropic priors, with a reference 
density which may be estimated by the algorithm itself. 
In the limit of zero noise, our approach is shown to reduce to  
the canonical MCE method generalized to a multi-period case. The method can
be used for a non-parametric pricing of American/Bermudan options with a 
possible weak path dependence.
}
\vspace{2.0cm}
\newcounter{helpfootnote}
\setcounter{helpfootnote}{\thefootnote} 
\renewcommand{\thefootnote}{\fnsymbol{footnote}}
\setcounter{footnote}{0}
\footnotetext{ Opinions expressed in this paper are those of the author, and
not necessarily of his employer. I want to thank E. Berger for useful 
discussions and reading the manuscript. All remaining errors are mine, of
course.
}
\renewcommand{\thefootnote}{\arabic{footnote}}
\setcounter{footnote}{\thehelpfootnote} 

                                                \end{titlepage}
\section{Introduction}

The linear inverse theory deals with the problem of reconstruction of a 
signal $ 
\{ f_i \} $ out of data (images) $ D = \{ d_1, \ldots , d_m \} $   
which are obtained from a signal by a linear operator $ K $, and may be 
corrupted with a noise $ \varepsilon $:
\beq
\label{inv_problem}
d_i = \sum_{j = 1}^{n} K_{ij} f_j + \varepsilon_i \; \; \; i = 1, \ldots, m
\eeq 
Since the number of available data $ m $ is typically much 
smaller than the dimension $ n $ of the probability space, 
inverse problems are highly underdetermined and ill-posed. 
Additional considerations, such as a prior knowledge and/or guiding 
regularization principles, become of a critical importance for solving such 
a problem.
Inverse problems are widely encountered in fields as diverse as image 
processing, geophysics, speech recognition,
computational linguistics,  molecular biology,
econometrics etc. Much progress in developing reliable and
robust methods of inversion was made in these areas over last decades, 
with mainstream efforts concentrated on Bayesian and 
Maximum Entropy (MaxEnt) or Minimum Cross Entropy (MCE) methods, see e.g. 
\cite{Jaynes,Brett,maxent_book,GCSR}. 

A popular paradigm in option pricing theory is the nonparametric option pricing
which is based on inferring a (risk-neutral) price probability density
out of market prices of liquid assets, and then calculating prices of other,
typically less liquid instruments as convolutions of payoff functions
with the inferred distribution.  
Depending on the data, a solution may not exist (that would signal 
arbitrage opportunities) or, more often, there may be multiple probability 
measures consistent with the same data (this means that the market is 
incomplete), thus the question of choosing the ``right'' solution arises. 
The problem of obtaining asset price distributions out of
option prices is thus a typical (linear) inverse problem, 
which can be addressed using well established methods of the 
inversion theory.

One popular method to find implied densities is to use
the MCE
inversion method as advocated and developed by Jaynes \cite{Jaynes},
without invoking Bayesian arguments. 
This approach was initiated in the financial literature by Gulko \cite{Gulko},
Buchen and Kelly \cite{BK}, and Stutzer \cite{Stutzer}. 
Within this method,  
the asset price distribution $ p(x) $ is found by requiring that 
it should minimize the relative entropy of $ p(x) $ and a given reference
distribution $ q(x) $, while reproducing the market prices of options in a 
chosen calibrating set (with all options having the same maturity). 
The relative entropy of two distributions $ p(x) $ and $ q(x) $ is 
given by the so-called Kullback-Leibler (KL) distance \cite{CT}
\beq
\label{KL_dis}
D[p(x)||q(x)] =  \int dx \, p(x) \log \frac{ p(x)}{q(x) }
\eeq
As $ \log(1/x) $ is a convex function, Jensen's inequality 
\beq
\label{Jensen} 
\int dy p(y) g[q(x)] \geq g \left( \int   dy \, p(y) q(y) \right)
\eeq
valid for any convex function $ g $ and probability 
$ p(y) $ such that $ \int dy p(y) = 1 $, shows that $ D[p(x)||q(x)] \geq 0 $
for any distributions $ p(x) , q(x) $. 

The reference density  $ q(x) $ in (\ref{KL_dis})
accumulates our prior knowledge and/or expectation
for the answer. 
There exist several possibilities to choose such a density. 
One of them is to take $ q(x) $ to be 
the historical asset price distribution, as suggested 
by Stutzer \cite{Stutzer}. One the other hand, if no historical information is 
available or if one believes that the risk-neutral distribution should be very
different from the historical one, one may take a lognormal
distribution or even a uniform reference 
distribution $ q(x) = const $. In the latter case, the KL distance becomes the 
(minus) absolute entropy of distribution $ p(x) $. The 
principle of maximizing 
the absolute entropy of a distribution (MaxEnt) 
is thus but a particular case of 
a more general principle of minimizing the relative KL entropy (MCE). 

With a uniform reference density $ q(x) = const $, 
the MaxEnt method gives the least informative and least 
prejudiced asset price distribution compatible with the market data. As was 
stressed by Gulko (who considered the case $ q(x) = const $ only), 
the MaxEnt principle thus formalizes and quantifies the 
Efficient Market Hypothesis in the form of a static characterization of 
market beliefs about future prices: the distribution must be such that it 
maximizes uncertainty about the future price movements.
By the meaning of entropy \cite{CT}, this means that the distribution should 
be of a highest possible absolute entropy, i.e. the lowest possible KL 
distance to a uniform distribution. 
The argument thus obviously generalizes 
to the case of a non-uniform density $ q(x) $ expressing the market belief
about the future: the most probable distribution $ p(x) $ should minimize 
the KL distance 
to the reference distribution $ q(x) $, while being consistent with the market
data.  
The MCE approach thus provides a consistent and theoretically sound
method of ``completing the market'', i.e. choosing the ``best'', maximally 
unbiased risk-neutral
probability measure if several possibilities 
compatible with the data are possible.
  
In the early works \cite{Gulko,BK,Stutzer},
the MCE method was applied to a simplest one-period model, with all options
in the calibrating set having the same maturity. Furthermore, no noise was 
supposed to be present in the data.
Later, the method was extended to a multi-period setting by Avellaneda and 
co-workers \cite{Avell} using methods of stochastic optimal control theory,
again without allowing for a noise in the data. For other recent 
applications of the canonical 
MCE approach, see Derman and Zou \cite{DZ} and Stutzer \cite{Stutzer2}.  

In reality, noise is always present in option prices, and can have a number 
of sources including non-synchronicity of trading, straight-out reporting 
errors, different trader perspectives of the market, etc. As was found  
by Buchen and Kelly \cite{BK} with Monte Carlo simulation, the MCE method 
often becomes 
unstable even for a modestly low level of noise in the data, leading 
sometimes to a fake multi-nodal form of an implied distribution. 
This behavior 
is traced back to the construction of the canonical MCE method (see below in
Sect.3) which enforces exact matching of the market prices and thus tends 
to overfit the data. These problems have led Buchen
and Kelly to the conclusion that the method should be abandoned altogether 
\cite{BK2} in favor of the so-called geophysical inverse theory based on a 
completely different approach. However, a way to overcome these difficulties 
within a natural generalization of the canonical entropy approch has been 
known for a long time to information theory researchers: It was stressed by 
Jaynes as early as in 1982 \cite{Jaynes82} that when the data
are noisy, the MaxEnt and MCE methods are unsuitable and should be 
substituted by the {\bf Bayesian inverse theory in combination with the 
MCE principle}. There exists an extensive literature 
\cite{Jaynes82,maxent_book} documenting both the need and the way to modify 
the canonical MCE method for dealing with noisy data. 

The contribution of the present work is two-fold. First, 
we develop a Bayesian
Inverse Theory approach to implied option pricing with noisy data, 
that naturally generalizes the canonical MCE method and reduces to the latter
in the limit of zero noise. The
only restriction we impose on the stochastic process for the asset price is 
the Markovian property. Beyond this, its nature (diffusion, jump-diffusion, 
etc.)
is left unspecified.
Second, we extend the method to the case of 
calibration to options of arbitrary maturities rather than one fixed 
maturity as in \cite{Gulko,BK,Stutzer}. In contrast to Avellaneda 
{\it et.al} \cite{Avell}, we build a recursive procedure.
We first calculate conditional transition 
probabilities with transition times being the adjacent maturities of options
in the calibrating set. This is performed recursively starting from a shortest 
maturity. When this is done, we may, if needed, fill the model 
``in between'' introducing
intermediate time levels between the shortest and the longest
maturities of the calibrating set. This may be necessary e.g. for pricing of 
a Bermudan option with exercise dates some (or all) of which 
do not belong in the given set of maturity dates.
Our method allows one to introduce an arbitrary number of intermediate
time levels and
calculate corresponding conditional transition densities.
The method involves two adjustable parameters related to 
modeling the noise and treating the reference density (see below) that may 
be either estimated from the data, or serve as a measure of the model risk.
As the output of our method is a set of 
conditional transition densities, it may be considered as 
a generalization of the implied tree methodology of Rubinstein \cite{Rub} and 
Derman and Kani \cite{DK} (see \cite{Jack} for a review). 
In this class of models, closest to ours is presumably a 
semirecombining implied tree construction of Brown and Toft \cite{BT}. 
Similarly to the method of \cite{BT}, we first find the conditional 
transition densities between different maturity dates in the calibrating 
set, and then calculate distributions for intermediate times of interest such 
as exercise dates for a Bermudan option. However,
in contrast to \cite{BT}, we do not discretize a stochastic
process for the underlying, do not introduce further time levels 
and, most important, use a different approach to fix the 
transition densities, which also allows one to account for the noise in data.
Computationally, our method is equivalent to performing a 
single MCE optimization for each maturity where data are available, and 
calculating  
four one-dimensional integrals for each additional time level. In 
addition, our algorithm is flexible in handling the reference density: at 
each step, it is generally given by a linear combination of  
a prior (e.g. historical) density and a conditional density 
of a previous step evolved to the current time (see Eq.(\ref{ref_dens}) 
below). The attractive feature of allowing for a non-vanishing value of 
$ 0 < \rho < 1 $ is that it smoothes the transition density 
in the time direction. 

Our presentation is organized as follows. In Sect.2 we introduce the basic 
recursive algorithm. The subsequent sections 3 to 7 
describe a single time step of the 
algorithm. In Sect.3 we outline the MCE approach to the problem valid when
no noise is present in the data. The Bayesian entropic inverse theory approach
to the problem needed when data are noisy is presented in Sect.4. In Sect.5 we 
explain how the noise variance should be integrated out in our Bayesian 
setting. A resulting non-linear optimization
problem with an objective function (which we call the effective 
Lagrangian using physics' nomenclature) is formulated in Sect.6, and 
its convexity properties are studied. The actual minimization of the 
effective Lagrangian is performed in Sect.7 using a linearization trick 
that leads to a drastic reduction in the computational cost. Sect.8 explains 
how to calculate transitions densities involving additional time levels. 
The final Sect.8 
contains our conclusion. 

\section{Recursive Algorithm}

We assume that we are given an ordered set of maturities 
$ \{ T_1, \ldots, T_n \} $.
For each maturity $ T_i $, $ i = 1, \ldots, n $, 
there are $ N_i $ available European option prices.
The option payoff functions are supposed to be linearly independent \cite{BK}.

Our scheme can be introduced as follows.
Let $ X_{1}, \ldots, X_{n} $ be the asset prices at times 
$ \{ T_1, \ldots, T_n \} $, respectively. We assume that a stochastic 
process governing the evolution of the asset price is Markovian.
Time today is $ t = 0 $. Consider now the time moment $ T_{i-1} $ for some 
$ i $.
We ask what the asset price distribution should be at 
time $ T_{i} $. Had we known the asset price $ X_{i-1} = x $ at time $ T_{i-1} $,  
following the MCE approach of \cite{Gulko,BK,Stutzer}
we would minimize the KL entropy for the 
distribution $ p(X_i| X_{i-1} = x_{i-1}) $. 
Since at most we might know only the distribution $ p(x_{i-1}) $ of the 
asset price at a future moment $ T_{i-1} $, the best we can do is to 
require that KL distance should be minimized on average. In other words,
we require minimization of the conditional KL entropy \cite{CT}
\beq
\label{cond_KL_recur}   
D[p(x_{i}|x_{i-1})||q (x_{i}|x_{i-1})] = \int dx  p(x_{i-1}) 
\int dy p(x_{i}|x_{i-1}) \log \left( \frac{ p(x_{i}|x_{i-1})}{ 
q(x_{i}|x_{i-1})} 
\right)
\eeq
with respect to $ p(x_{i}|x_{i-1}) $ for each $ i = 1, \ldots, n $, 
subject to the condition of matching the observed market option prices. 
This suggests a forward recursive algorithm for solving the problem.
(The matching of data can be done either in the form of 
absolute constraints or in the Bayesian framework 
via a likelihood function. As will be clear below, the latter
preserves the forward
structure of the relations between different distributions.)  

It should be noted that this recursive algorithm is 
equivalent to a direct minimization of the KL entropy of the joint 
distribution $ p(X_{1}, \ldots, X_{n}) $ 
\beq
\label{joint_KL}
D[ p(x_1, \ldots, x_n) ||  q(x_1, \ldots, x_n) ] = \int dx_1 \ldots dx_n \, 
p(x_1, \ldots, x_n) \log \frac{ p(x_1, \ldots, x_n)}{q(x_1, \ldots, x_n)}
\eeq
Indeed, using the 
chain rule for relative entropy \cite{CT}, we obtain
\bea
\label{chain_rule}
D[ p(x_1, \ldots, x_n) ||  q(x_1, \ldots, x_n) ] &=& \sum_{i=1}^{n}
D[ p(x_i|x_{i-1}, \ldots, x_{1}) || q(x_i|x_{i-1}, \ldots, x_{1}) ]
\nonumber \\
&=& 
\sum_{i=1}^{n} D[ p(x_i|x_{i-1}) || q(x_i|x_{i-1}) ]
\eea
where on the last step we used the Markov property $ 
p(X_i|X_{i-1}, \ldots, X_{1}) = p(X_i|X_{i-1}) $ (and analogously for 
the reference distribution). As each term in the series 
(\ref{chain_rule}) is non-negative, minimization of the 
whole expression 
amounts to a sequential minimization of each successive term in the 
series\footnote{This statement is valid assuming that a minimum is unique.
As will be argued below in Sect. 3, in our optimization problem the minimum,
if exists, is unique.}.
This is a special property of a Markov process which enables one to 
use a forward calculation to calculate probabilities\footnote{A similar 
forward 
procedure is used for implied trees in the approaches of 
\cite{Rub,DK,BT}.}.
The two schemes produce identical results, as can also be 
checked in a simple two-period setting. We note that it is the scheme of 
minimization of the joint KL distance (\ref{joint_KL})
that was adopted by Avellaneda
{\it et.al} \cite{Avell}. 
We prefer the former forward recursive scheme because it is easier to 
implement, and also because the conditional transition probabilities
needed for pricing other securities are its direct outputs, while in the 
scheme adopted in \cite{Avell} their calculation includes additional 
integrations. Last but not least, a generalization to the case of noisy
data is quite straightforward in the recursive scheme but not in the method
of \cite{Avell}. 
   
Our recursive algorithm proceeds as follows. 
Suppose that the problem is solved up to time $ T_k $, i.e. we know all
conditional transition probabilities $ p(y, T_{i+1} | x , T_{i} ) $ for 
$ i < k $ as well as the marginal density  (conditional on the initial
price) $ p(x) \equiv p(x,T_k | x_0, 0) $  where time today is $ t = 0 $ and
todays asset's value is $ x_0 $.
All reference conditional transition densities are calculated as well.
The recursion step is then performed as follows. \\
{\bf 1}. Set the transition times 
$ t \equiv T_k $ , $ T \equiv T_{k+1} $. \\
{\bf 2}. Calculate the conditional transition probability $ p(y|x) \equiv 
p(y,T| x,t) $ of the asset price distribution at time $ T $ conditional
on its value at time $ t $. The calculation is performed using the Bayesian 
inverse theory, and requires the knowledge of the marginal density $ p(x) $ at
time $ t $. It also involves the reference conditional transition density
$ q(y|x) $ for this step.\\
{\bf 3}. Calculate the reference conditional transition density for 
the next step.
To this end, we shift all time arguments (which 
enter the calculation through options payoff functions) in 
$ p(y,T_{k+1}| x,T_k) $
to the
next time level $ (T_k , T_{k+1}) \rightarrow (T_{k+1}, T_{k+2}) $. 
Call this density $ q_{s} (y|x) $. A 
general form for the reference density for the next step thus would be 
\beq
\label{ref_dens}
q(y|x) = ( 1 - \rho) \, q_{h} (y|x) + \rho \,  q_s (y|x)
\eeq
where $ q_{h} (y|x) $ stands for the conditional density from time 
$ T_{k+1}$ to time $ T_{k+2} $ calculated with a  prior model (or a 
historical price distribution), and $ 0 \leq \rho \leq 1 $ is a free parameter
of the model. The limit $ \rho \rightarrow 1 $ corresponds to maximum 
smoothing  of a conditional transition density in the time 
direction\footnote{ The choice of $ q_s (y|x) $ as a prior distribution is 
reasonable if the real process is stationary, which means that the 
conditional density $ p(y,T| x,t) $ depends only on the difference $ T - t $.
We can further neglect an implicit dependence of 
Lagrange multipliers on $ T- t $, which is a good zero approximation so long 
as  $T_{k+2} - 
T_{k+1} $ is close (or equal) to $ T_{k+1} - T_{k} $.} \\
{\bf 4}. Calculate a new marginal density 
\[
\int dx  \,p(x)  p(y|x) = \int dx  \,p(x|x_0)  p(y|x)
\]
to be used to find the conditional transition density on the next step. \\
{\bf 5}. Increment $ k \rightarrow k + 1 $. \\
{\bf 6}. If $ k < n - 1 $, go to step 1. \\
{\bf 7}. If needed, calculate conditional transition densities involving additional time 
levels.  

In what follows, Sects. 3 to 7 describe step 2 in the above scheme, while
Sect.8 explains the calculation on step 7. 
 
\section{Canonical Minimum Cross Entropy Method}

Assuming that the unconditional density $ p(x) $ at time $ t = T_k $ is known,
we want to find the transition probability density $ p(y| x) $ for 
transition to the interval $ [ y, y + dy] $ at time $ T \equiv T_{k+1} $ 
conditional
on being in state $ x $ at time $ T_i $. We require that  $ p(y| x) $ should 
be close to a prior density $ q(y|x) $, i.e. should minimize the 
conditional KL entropy functional
\beq
\label{KL}
D[p(y|x)||q (y|x)] = \int dx  p(x) 
\int dy p(y|x) \log  \frac{ p(y|x)}{ q(y|x)} 
\eeq
subject to all constraints imposed on  $ p(y| x) $. The first constraint
is the normalization equation 
\beq
\label{norm}
\int \, dy p(y|x) = 1 
\eeq
that should be imposed for all values of $ x $. 
Further constraints are imposed by 
matching observed option prices. We consider a set of $ N \equiv N_{k+1}$ 
European
options maturing at time $ T = T_{k+1} $, with $ \{ f_i (x,y) \}$ being the 
payoff functions. For the sake of generality, we   
retain the $ x $ argument in  $ f_i (x,y) $, which may 
account for a certain path dependence. More common, however, is calibration 
to European vanilla call and put options whose payoff functions $ f_i (x,y)
= f_i (y) $  are 
\bea
\label{payoffs}
f_i (y) &=& D(T -t) \, {\mbox max}( y - K_i, 0) \; \; \; (call) \\
f_i (y) &=& D(T -t) \, {\mbox max}( K_i - y, 0) \; \; \; (put)
\eea
where $ \{ K_i \} $ are strikes and $ D(T -t) $ are (deterministic) discount 
factors: $ D(T-t) = \exp[ -r (T-t)] $.
An exact matching of market prices imposes the constraints
\beq
\label{constr}
\int dx p(x) \int dy p(y|x) f_i (x,y) = C_i \; \; \; , \; \; \; i = 1,\ldots,
	N
\eeq
where $  C_i $  stands for a (noiseless) price of the $i$th option.

The problem of minimization of (\ref{KL}) subject to constraints (\ref{norm}) 
and (\ref{constr}) is a standard problem of constrained optimization
that is solved by the Largange multiplier method. To satisfy the normalization
constraint (\ref{norm}) for each value of $ x $, one needs a  
Lagrange multiplier function $ \xi(x) p(x) $. Here 
the multiplicative factor $ p(x) $ (known from the previous step) 
is introduced for a later convenience.
Constraints (\ref{constr}) are imposed using a discrete set of $ N 
 $ Lagrange multipliers $ \{ \lambda_i \} $. We thus  
arrive at the following Lagrangian function
\bea
\label{Lagrangian}
L &=&  \int dx p(x) \int dy p(y|x) \log  \frac{ p(y|x)}{ q(y|x)} 
 - \int dx \xi (x) p(x) \left( \int dy p(y|x) - 1 \right)  \nonumber \\
&-& \sum_{i} \lambda_i \left( \int dx p(x) \int dy p(y|x) f_i (x,y) - 
C_i \right)
\eea
The Lagrangian (\ref{Lagrangian}) is minimized when its total variation 
vanishes, i.e.
\beq
\label{extremals}
\frac{ \delta L}{ \delta p(y|x) } = 0 \; \; , \; \; 
\frac{  \delta L}{ \delta \xi (x)} = 0 \; \; , \; \; 
\frac{  \partial L}{  \partial \lambda_i}  = 0 \; , \; i = 1, \ldots, N
\eeq
The solution of the first two equations in (\ref{extremals}) reads
\beq
\label{canon_dist}
p(y|x) = \frac{q(y|x)}{Z_{\lambda}(x)}
	  \exp \left( \sum_{i} \lambda_i  f_i (x,y)\right)
\eeq
where 
\beq
\label{z_factor}
Z_{\lambda}(x) = \int dy \, q(y|x) \exp \left( 
 \sum_{i} \lambda_i  f_i (x,y)\right)
\eeq
What remains is to eliminate the Lagrange multipliers $ \lambda_i $ by substituting the solution (\ref{canon_dist}) in  
the last of Eqs.(\ref{extremals}). This results in a set of nonlinear 
equations
\beq
\label{NR_problem}
\int \int p(x) \frac{q(y|x)}{Z_{\lambda}(x)} 
\exp \left( \sum_{i} \lambda_i  f_i (x,y)\right)
f_i (x,y) = C_i  \; \; \; , \; \; \; i = 1, \ldots, N
\eeq
This problem can be solved e.g. by a multidimensional Newton-Raphson procedure
\cite{AAD}. As shown in \cite{BK}, the 
Jacobian matrix in this problem is 
given by the covariance matrix of the constraint functions
\beq
\label{cov_matr}
J_{ij} = {\mbox cov} \left[ f_i (x,y), f_j (x,y) \right]
\eeq
and thus is positive definite for linearly independent constraints, 
which is a necessary condition for the 
Newton-Raphson method to be applicable. In addition,
$ J_{ij} $ should be well-conditioned to ensure stability of the algorithm.
As  discussed in \cite{BK}, the 
Jacobian may become ill-conditioned if market prices 
are noisy. In case a case, a numerical solution for the Lagrange multipliers
may become unstable.
	
However, market prices are always noisy. As the classical MCE method 
enforces the exact matching of constraints (\ref{constr}), it 
tends to overfit the data. As a result, implied MCE probability 
distributions often become unstable and juggled. Intuitively, we might 
want to substitute the exact constraints  (\ref{constr}) by approximate 
ones that would smooth out the noise in the data. It will be shown in the next 
sections that the Bayesian inverse theory with entropic priors provides 
just such a sort of modification of the MCE method.

\section{Noisy Data: Bayesian Entropic Inverse Theory} 

In Bayesian probability theory \cite{Jaynes}, 
the posterior probability 
$ P(M|D,I) $ of a model (probability density, in our case) 
$ M $ given the data $ D $ and information $ I $ 
is determined by the Bayes formula 
\beq
\label{Bayes}
P(M|D,I) = P(M|I) \, \frac{P(D|M,I)}{P(D|I)}
\eeq
where $ P(M|I) $ the prior probability of the model assigned before the 
data is observed, and $ P(D|M,I) $ is the likelihood function measuring 
probability to observe the data $ D $ given the model $ M $. Finally, the 
so-called evidence
$ {P(D|I)} $ is the total probability of observing the data given the 
information $ I $. As it does not depend on the model $ M $ whose probability 
we want to calculate, it merely serves as an irrelevant normalization factor
for the posterior probability $  P(M|D,I) $. 

To solve the inverse problem of finding the transition probabilities
based on data and a prior knowledge, we apply the Maximum Aposteriori
Probability (MAP) approach \cite{Jaynes,maxent_book}. According to this method, the 
most probable 
model is one that maximizes the posterior probability 
$ P(M|D,I) $ in (\ref{Bayes}) i.e. the product of 
the prior $ P(M|I) $  and the likelihood function $ P(D|M,I) $.
Note here the difference with the usual Maximum Likelihood method which 
requires fixing the prior in advance and optimizing the likelihood function
alone. On the contrary, the MAP method allows the prior to carry 
unknown parameters and/or functional dependence. 

We start with assigning the prior probability  $ P(M|I) $.
The following combinatorial argument \cite{Rodr} that extends one 
suggested by Jaynes \cite{Jaynes}, is of use here (see also \cite{Caticha} for 
a recent
discussion of entropic priors). For simplicity, we will
first consider marginal, not conditional probabilities, and discretize  
the state space $ X = \{ 1, 2, \ldots , k \} $. Let $ q_j $ be the probability
of the $ j$th element of $ X $. We can imagine an urn with known proportions
$  q_j $ of balls of type $ j$, for $  j = 1, 2, \ldots , k$. If we randomly 
draw $ n $ balls from the urn, then the probability of observing $ n_i = 
p_i n $ balls of each class $ i $ is given by the multinomial distribution
\beq
\label{multinom}
P(p|q) = P(n_1, \ldots, n_k | q) = \frac{ n!}{n_1 ! n_2 ! \ldots n_k !} \, 
q_1^{n_1} \ldots q_{k}^{n_k} 
\eeq
which is the probability of seeing a $p$-distribution when sampling $n$ times
from a $q$-distribution. We assume that $ n \gg 1 $.
Taking the logarithm of both sides of 
(\ref{multinom}) and using the Stirling formula
\beq
\label{Stir}
\log m! = m \, \log m - m + o(m)
\eeq
we obtain 
\beq
\label{ent_pr}
P(p|q) \propto \exp \left( - n \sum_{j =1}^{k} p_j \log \frac{p_j}{q_j} \right)
\eeq
Note that for the case of a uniform reference
distribution $ q_j = q = const $, this line of reasoning boils down to 
Jaynes' definition of the prior
\beq
\label{Jaynes_pr}
P(p) \propto \exp \left( - \sum_{j =1}^{k} p_j \log p_j \right) 
\eeq
while the exact value of parameter $ n $ becomes irrelevant in this limit, as 
will be clear below. 
Extending the argument to the case of conditional 
probabilities, we write down the prior probability of distribution 
$ p(y|x) $ in our problem:
\beq
\label{my_prior}
P[p||q] = \exp \left[ - h \int dx  p(x) \int dy p(y|x) 
\log  \frac{ p(y|x)}{ q(y|x)} 
 \right] \prod_{x} \delta \left( \int dy \, p(y|x) - 1 \right)
\eeq
where we added the $ \delta $-function constraint for each value of $ x $
to ensure the correct normalization. Here $ h \gg 1 $ is a
parameter controlling the relative weight of the entropy term in the 
posterior probability.

Our next task is to compute the likelihood function $ P(D|M,I) $. To this
end, we assume that noise in prices is Gaussian\footnote{Note that the 
assumption of an additive noise in the prices is, strictly 
speaking, theoretically problematic, as potentially it allows for negative 
prices. However, assuming that a noise is not too large, a probability of
this to occur will be negligible. Yet, presumably a more realistic 
approach would be to model an additive noise for logarithms of option prices. 
We will leave a study of such a model for a future work.}:
\beq
\label{noisy_constr}
\int dx p(x) \int dy p(y|x) f_i (x,y) = C_i + \varepsilon_i 
\; \; \; , \; \; \; i = 1,\ldots, N  
\eeq
where 
\beq
\label{noise}
{\bf E}[\varepsilon_i ] = 0 \; \; \; , \; \; \; 
{\bf E}[\varepsilon_i^2 ] = \sigma_i^2 
\eeq
The likelihood function is therefore given by the following expression:
\beq
\label{likelihood}
P \left[ D| p, \sigma \right] = 
\frac{1}{ (2 \pi)^{N/2} \prod_{i} \sigma_i }
\exp \left[ - \sum_{i} \frac{1}{2 \sigma_i^2} \left( \int \int dx dy p(x)
p(y|x) f_i(x,y) - C_i \right)^2 \right]  
\eeq 
If the prior distribution for the noise variance is 
given by some function $ P(\sigma) $, the full prior in our problem 
becomes \footnote{A fully Bayesian approach would also require a specification
of a prior for parameter $ h $ in (\ref{my_prior}). We will not attempt 
doing this, but will rather view $ h $ as a fixed parameter, and will study
stability of our results with respect to variations of $ h$.} 
\beq
\label{full_prior}
P(M|I) = P(\sigma) \, P[p||q]
\eeq
where $  P[p||q] $ is given by (\ref{my_prior}).
As we are not interested in estimating the noise variances $ \sigma_i^2 $, they
represent the so-called nuisance parameters (hyperparameters) 
that should be integrated out
according to the Bayesian approach \cite{Jaynes,Brett}. Therefore, for the 
posterior probability we obtain
\beq
\label{posterior}
P\left[ p | D \right] \propto P\left[ p||q \right] \,
 \int_{0}^{\infty} \prod_{i} d \sigma_i \, P( \sigma) 
P \left[ D| p, \sigma \right]
\eeq
where $  P\left[ p||q \right]  $ and $ P \left[ D| p, \sigma \right] $
are defined in  (\ref{my_prior}) and (\ref{likelihood}), respectively.
According to the MAP approach, this final expression should be maximized with
respect to $ p(y|x) $ in order to find the most probable transition probability
$ p(y|x) $ compatible with the data while remaining as close as possible to 
the reference density $ q(y|x) $.

Eq.(\ref{posterior}) allows one to readily see how the usual MCE method
is reproduced in the limit of zero noise. This limit corresponds to  
knowledge of the prior for noise with certainty:
\beq
\label{delta_np}
P(\sigma) = \prod_{i} \delta \left( \sigma_i - \sigma_{i}^{(0)} \right) 
\eeq
and taking the limit $ \sigma_{i}^{(0)} \rightarrow 0 $  after 
the integral in (\ref{posterior}) is calculated. This yields
\beq
\label{back_to_ME}
P\left[ p | D \right] \propto P\left[ p|q \right]  \prod_{i} 
\delta \left( \int \int dx dy p(x)
p(y|x) f_i(x,y) - C_i \right) 
\eeq
Using the integral representation of the $ \delta $-function
\beq
\label{int_delta}
\delta(x) = \frac{1}{2\pi} \int_{- \infty}^{ \infty} d \omega \exp( i \omega x)
= \frac{1}{2\pi i} \int_{- i \infty}^{ i \infty} d \lambda \exp( - \lambda x)
\eeq
we see that in the limit of zero noise the MAP approach exactly reproduces
the MCE method, with the saddle point values of parameters $ \lambda $ 
in (\ref{int_delta}) for all $ \delta $-functions in (\ref{back_to_ME})
being exactly equivalent to Lagrange multipliers in (\ref{Lagrangian}).
It is easy to see that the exact value of parameter $ h $ becomes irrelevant 
in this limit, as it can be eliminated by a rescaling of Lagrange multipliers.
\section{Integrating out Noise Variance}

In general, a noise in option prices does persist. To integrate it out
according to the Bayesian theory, we should define a model for the prior 
distribution $ P(\sigma) $. Given the prior, we then calculate the 
integral
\beq
\label{integral}
\Omega[p] = \int_{0}^{\infty} \prod_{i} d \sigma_i P(\sigma) 
P \left[ D| p, \sigma \right] 
\eeq
One possibility is to assume the $ \delta $-function prior (\ref{delta_np})
but with $ \{ \sigma_{i}^{(0)} \} $ fixed and positive, assumed to 
be known numbers. In this case the integral (\ref{integral}) is trivially 
calculated with the result $ P \left[ D| p, \sigma^{(0)} \right] $. This 
is in spirit of the frequentist approach to statistics where the noise 
variance is a fixed albeit unknown number which can
be estimated from the data.

Another possibility is to choose a smooth continuous 
prior for the noise variance. 
If no information about the noise variance is 
available, one choice would be to introduce a completely non-informative
prior $ P(\sigma) = const $. As was noticed by Jaynes (see \cite{Jaynes})
this choice is unsatisfactory on the grounds that it does not respect the 
invariance principle: if we change the variable $ \sigma \rightarrow \sigma' $,
the prior $ P(\sigma) $ will generally change. By requiring that the prior
$  P(\sigma) $ should be invariant with respect to a change of scale
\beq
\label{scale} 
\sigma^{\star} = a \sigma \; \; \; , \; \; \; 
P(\sigma^{\star}) = J^{-1} P(\sigma)
\eeq
where in our case the Jacobian $ J = a $, we obtain the solution
\beq
\label{Jeffrey_prior}
P(\sigma) = \frac{const}{ \sigma}
\eeq
This is called Jeffrey's prior. It is of frequent use in Bayesian approaches.
Although the prior (\ref{Jeffrey_prior}) is unnormalizable (so-called improper
prior), its substitution in (\ref{integral}) leads to a converging integral.
Another approach considered 
in the Bayesian literature is to choose prior $ P(x) $ 
distributions for which a posterior distribution $ P(x|D) $ can be easily 
calculated. The most popular choice are the so-called
conjugate priors \cite{GCSR}. They possess the intuitive feature that 
starting with a certain functional form for the prior, one ends up with 
a posterior of the same functional form, but with parameters updated by the 
sample information. For a Gaussian noise, the conjugate prior is therefore
given by the Gamma distribution in $ 1/ \sigma^{-2} $ :
\beq
\label{Conj_prior}
P(\sigma) \propto 
(\sigma^2)^{1 - \alpha } \, 
\exp \left( - \beta \sigma^{-2} \right) 
\eeq     
Here $ \alpha >  0 $ and $ \beta > 0 $. 
An improper prior corresponds to the limit 
$ \alpha \leq 3/2 $ and $ \beta \rightarrow 0 $. 
Note also that the particular choice $ \alpha = \nu/2 \, , \; \beta = 1/2 $ 
corresponds to the Chi-square distribution with $ \nu $ degrees of freedom
\footnote{
This may hint on a bridge to the classical approach to statistics whose 
fundamental theorem states that if $ S^2 $ is the variance of a random sample 
of size $ n $ from a normal distribution with the mean $ \mu $ and variance 
$ \sigma^2 $, then the random variable $ (n-1) S^2 / \sigma^2 $ has a 
Chi-square distribution with $ (n-1) $ degrees of freedom.}.

In what follows we analyze 
both $ \delta $-function  (\ref{delta_np}) and  
conjugate (\ref{Conj_prior}) choices for the prior.
Jeffrey's prior is recovered by the particular choice
\beq
\label{Jeffrey_choice}
\alpha = \frac{3}{2} \; \; \; , \; \; \; \beta = 0  
\eeq
in the conjugate prior (\ref{Conj_prior}). (As will be seen in the next
section,this choice may however be undesirable as it does not lead to 
a convex optimization problem.) 
We also note that a  
conjugate prior $ p (x) \propto x^{\alpha -1}e^{- \beta x} $  
can be viewed as the MCE prior maximizing entropy given
integral constraints $ E[X] = g_1 \; , \; E[ \log(X) ] = g_2 $. Recall   
that the entropy itself can be interpreted as an expected value $ H(X) = 
E[ - \log \, p(X) ] $, so that the second constraint here actually fixes the 
entropy of the distribution.

\section{Effective Lagrangian}

If we choose a conjugate prior (\ref{Conj_prior}),
integral (\ref{integral}) is calculated  
using the well-known formula for the Gamma distribution
\beq
\label{int_gamma}
\int_{0}^{\infty} dy \, y^{b-1} e^{-a y} = \Gamma(b)\, a^{-b}
\eeq
and assuming identical distributions 
for all $ \sigma_i $. 
It is convenient here to define the ``effective potential'' 
\beq
\label{eff_potent}
\Phi[p] =  - \log \Omega [p] 
\eeq
For the conjugate prior (\ref{Conj_prior}) we thus obtain
\beq
\label{Conj_post}
\Phi_{CP}[p] = (\alpha  - 1) \sum_{i = 1}^{N} \log \left[  \left( \int \int 
dx dy p(x) p(y|x) f_i (x,y) - C_i \right)^2 + 2 \beta \right]
\eeq
plus an inessential constant.
If instead the $ \delta $-function 
prior (\ref{delta_np}) is used, the effective potential is (hereafter we omit
the subscript $(0)$ from $ \sigma_i^{(0)} $) 
\beq
\label{delta_post}
\Phi_{DF}[p] =  
  \sum_{i} \frac{1}{2 \sigma_i^2} \left( \int \int dx dy p(x)
p(y|x) f_i(x,y) - C_i \right)^2   
\eeq
Using (\ref{delta_post}), 
(\ref{Conj_post}), (\ref{my_prior}) and (\ref{posterior}), we find that
the MAP probability density $ p(y|x) $ is 
\beq
\label{MAP_sol}
p(y|x) = \arg \min_{p(y|x)} \max_{\xi(x)} L_{eff}[p,\xi]
\eeq 
with the following ``effective Lagrangian'':  
\bea
\label{eff_Lagrangian}
L_{eff}[p,\xi ] &=& h \int dx p(x) \int dy p(y|x) \log \left( \frac{ p(y|x)}{ q(y|x)} 
\right) \nonumber \\
&-& \int dx \xi (x) p(x) \left( \int dy p(y|x) - 1 \right)  
+ \Phi[p]
\eea
where $ \Phi[p] $ is defined by either (\ref{Conj_post}) or (\ref{delta_post}).
 
To find whether the resulting Lagrangian admits a unique minimum or multiple 
local minima,
we study its convexity properties. 
If the 
effective Lagrangian (\ref{eff_Lagrangian}) is convex, a solution of the 
minimization problem, if exists, is unique. Recall that a 
functional $ F[f(x)]  $ is convex if 
\beq
\label{convexity}
F[ \lambda f_1(x) + \bar{\lambda} f_2 (x) ] \leq  \lambda \, F[f_1(x)] + 
\bar{\lambda} \, F[ f_2 (x) ]  
\eeq
for any distributions $ f_1(x), f_2(x) $ and $ 0 \leq \lambda \leq 1 $, $ 
\bar{ \lambda} \equiv 1 - \lambda $.

Convexity of the first term in (\ref{eff_Lagrangian}) is easy to establish:
\bea
\label{first_conv}
 & & D \left[ \lambda p_1(y|x) + \bar{\lambda}  p_2(y|x) || q \right] - 
 \lambda \, D \left[ p_1(y|x)|| q \right] - \bar{\lambda} \, 
 D \left[ p_2(y|x)|| q \right] \nonumber \\ 
& &  = \lambda \, \int \int dx dy p(x) p_1(y|x)
\log \left( \frac{ \lambda  p_1(y|x) + \bar{\lambda} p_2(y|x)}{  p_1(y|x)}
\right) \\  
& & + 
\bar{\lambda} \,  \int \int dx dy p(x) p_2(y|x)
\log \left( \frac{ \lambda  p_1(y|x) + \bar{\lambda} p_2(y|x) }{ p_2(y|x)}
\right) \leq 0 \nonumber
\eea
because both terms are non-positive by Jensen's inequality (\ref{Jensen}).

Consider now the effective potential term in (\ref{eff_Lagrangian}).
Convexity is easy to check for the $ \delta $-function prior 
(\ref{delta_post}) for which
\bea
\label{sec_conv}
& & \Phi_{DF}[\lambda p_1 + \bar{\lambda}  p_2 ] - \lambda \, \Phi{DF}[ p_1] - 
\bar{\lambda} \, \Phi_{DF}[ p_2 ] \nonumber \\
& & = - \, \frac{ \lambda \bar{\lambda}}{ \sigma_0^2} \left[ \int \int dx p(x)
\left( p_1(y|x) - p_2(y|x) \right) f_i (x,y) \right]^2 \leq 0
\eea 
Consider now the effective potential obtained with the conjugate prior 
(\ref{Conj_post}).
 Since $ - \log(y) $ is 
a convex function, convexity of the effective potential (\ref{Conj_post})
is guaranteed if $ 0 < \alpha < 1 $ and $ \beta = 0 $. For other values of 
parameters $ \alpha, \beta $, convexity is lost. 
In particular, the choice of Jeffrey's prior (\ref{Jeffrey_choice})
does not lead to a convex optimization problem. 
This does not necessarily mean that it cannot give sensible 
results, but this case is bound to be more complicated than a convex problem, 
and will not be analyzed here. 
In what follows, 
the conjugated prior (\ref{Conj_prior}) will therefore be  
considered only in the parameter range $ 0 < \alpha < 1 $ , $ \beta = 0 $.

\section{Minimization of Effective Lagrangian}

As it stands, the effective Lagrangian (\ref{eff_Lagrangian}) 
appears to be considerably more 
complicated than the corresponding Lagrangian (\ref{Lagrangian}) of the 
canonical MCE method applicable when no noise is present. The difference
is due to a non-linearity of the effective potential in 
(\ref{eff_Lagrangian}), which is in contrast to a corresponding linear term
in (\ref{Lagrangian}). The fact that this potential term is linear enables 
an analytical solution (\ref{canon_dist}) of the 
variational problem in the classical MCE method, while 
a numerical scheme is needed only to fix the 
values of the Lagrange multipliers. On the contrary, in the present method
there are no independent Lagrange multipliers, but the problem seems to 
boil down to a  highly non-linear integral equation due to a non-linearity 
of the effective potential.

With the MAP approach, a simple trick may be suggested that leads to a drastic
simplification of the problem. The trick is akin to one used to 
treat non-linearities in quantum field theory. We propose to linearize the 
effective potential in (\ref{eff_Lagrangian}) by introducing new variables 
$ U_i , \lambda_i $ and minimizing the following Lagrangian\footnote{Here we
rescale all terms by $ 1/h $.} 
\bea
\label{ext_Lagrangian}
& & L_{ext}[p,\xi, U , \lambda ] =  \int dx p(x) \int dy p(y|x) \log \frac{ p(y|x)}{ q(y|x)} 
- \int dx \xi (x) p(x) \left( \int dy p(y|x) - 1 \right) \nonumber \\  
& & + \Phi[U] + \sum_{i} \lambda_i \left( U_i - \int \int dx dy \, p(x) 
p(y|x) f_i (x,y) +  C_i \right)  
\eea
where $ \Phi[U] $ stands for the effective potential  (\ref{delta_post}) or 
(\ref{Conj_post}) in which all combinations $ \int \int dx dy \, p(x) 
p(y|x) f_i (x,y) -  C_i $ are substituted by $ U_i $:
\beq
\label{phi_u}
\Phi_{DF}[U] =   \sum_{i} \frac{1}{2 h \sigma_i^2}U_i^2 \; \; , \; \; 
\Phi_{CP}[U] =  - \frac{1 - \alpha}{h} \sum_{i = 1}^{N} \log \left[ U_i^2  \right]
\eeq 
 It is easy to see that
we come back to the original effective Lagrangian (\ref{eff_Lagrangian}) if 
we  first minimize (\ref{ext_Lagrangian}) over $ U_i $ and $ \lambda_i $ 
with $ p(y|x), \xi $ fixed 
and substitute the result back into (\ref{ext_Lagrangian}).
The MAP solution $ p(y|x) $ is therefore  determined by the following set of
minimization equations:
\beq
\label{minim_ext}
\frac{\delta L_{ext}}{ \delta p(y|x) } = 0 \; \; , \; \; 
\frac{\delta L_{ext} }{ \delta \xi(x) } = 0 \; \; , \; \; 
\frac{\partial L_{ext} }{ \partial U_i} = 0 \; \; , \; \; 
\frac{\partial L_{ext} }{ \partial \lambda_i} = 0 
\eeq
First two equations in (\ref{minim_ext}) yield 
\beq
\label{canon_dist_ext}
p_{\lambda}(y|x) = \frac{q(y|x)}{Z_{\lambda}(x)}
	  \exp \left( \sum_{i} \lambda_i  f_i (x,y)\right)
\eeq
where 
\beq
\label{z_factor_ext}
Z_{\lambda}(x) = \int dy \, q(y|x) \exp \left( 
 \sum_{i} \lambda_i  f_i (x,y)\right)
\eeq
Up to this point the solution is identical to one given by the canonical 
MCE method. The difference arises when it comes to a solution for 
the Lagrange multipliers $ \lambda_i $. Using (\ref{canon_dist_ext}) and 
the last two of equations (\ref{minim_ext}), we end up with a set of 
non-linear equations for $ \lambda_i $:
\beq
\label{NR_problem_ext}
\int \int dx dy \, p(x) \frac{q(y|x)}{Z_{\lambda}(x)} 
\exp \left( \sum_{i} \lambda_i  f_i (x,y)\right)
f_i (x,y) = C_i  +  U_{i}^{\star}(\lambda_i) \; \; \; , 
\; \; \; i = 1, \ldots, N
\eeq
where $ U_{i}^{\star}(\lambda_i) $ stands for a solution of the equation
\beq
\label{ustar}
\frac{ \partial \Phi(U)}{ \partial U_{i} } = - \lambda_i 
\eeq
Equation (\ref{ustar}) yields
\bea
\label{solut}
U_{i}^{\star}(\lambda_i) & = - h \sigma_i^2 \lambda_i   & {\mbox (DF)} 
\nonumber \\
U_{i}^{\star}(\lambda_i) & = \frac{ 2 (1 - \alpha) }{ h \lambda_i}   
& {\mbox (CP)}
\eea
for the effective potentials (\ref{phi_u}), respectively. 
As one can see, for the choice $ \Phi(U) = \Phi_{DF}(U) $ the solution
smoothly goes to the canonical MCE solution in the limit $ \sigma_i 
\rightarrow 0 $. 

Let us now analyze convexity properties of the resulting problem 
of optimization in parameters $ \{ \lambda_i \} $. By the back 
substitution of the MAP distribution (\ref{canon_dist_ext}) and solution
(\ref{solut}) of  (\ref{ustar}) into   
(\ref{ext_Lagrangian}), we arrive at the following function 
\beq
\label{ul_opt}
F(\lambda) \equiv  \min_{p(y|x), \xi, U} \; L_{ext}[p, \xi, U, \lambda] 
= - \int dx \, p(x) \log Z_{\lambda}(x)  + \Psi(\lambda)
\eeq
where 
\beq
\label{psi}
\Psi(\lambda) = \Phi(U^{\star}) + \sum_{i} \lambda_i \left( 
U_i^{\star} + C_i \right)
\eeq
and  $ U_{i}^{\star}(\lambda_i) $ is determined by Eq.(\ref{ustar}).
In particular, for the $ \delta $-function and conjugate priors we obtain
\bea
\label{psi_factors}
\Psi_{DF}( \lambda) &=&  \sum_{i} \left( 
\lambda_i C_i - \frac{ h \sigma_i^2}{2} \lambda_i^2  \right) \; \; , 
\; \; \frac{ \partial^2 \Psi_{DF} }{
\partial \lambda_i \partial \lambda_j } \leq 0 \nonumber \\
\Psi_{CP}( \lambda) &=& \frac{1 - \alpha}{h} \sum_{i} \log \lambda_i^2   
+ \sum_{i} \lambda_i C_i + const \; \; , 
\; \; \frac{ \partial^2 \Psi_{CP} }{
\partial \lambda_i \partial \lambda_j } \leq 0 
\eea
For the Hessian matrix  
we find
\bea
\label{sec_der_lambda}
\frac{ \partial F^2}{ \partial \lambda_i 
\partial \lambda_j } &=& - \int dx p(x) \left[
\int dy p_{\lambda}(y|x) f_i (y) f_j (y) - \int dy p_{\lambda}(y|x) 
f_i (y) \int dy' p_{\lambda}(y'|x) 
f_j (y') \right]  \nonumber \\ 
& + & \frac{ \partial^2 \Psi }{
\partial \lambda_i \partial \lambda_j } \equiv 
  - \int dx p(x) \, {\mbox cov} (f_i, f_j) (x)  + \frac{ \partial^2 \Psi }{
\partial \lambda_i \partial \lambda_j }
\leq 0  
\eea
which is because a covariance matrix of constraints is positive definite for 
linearly 
independent constraints. 
We conclude that the function $ F(\lambda) $ (\ref{ul_opt}) is 
concave for both 
choices of the effective potential, and therefore the solution of the 
linearized problem (\ref{ext_Lagrangian}), if exists, is unique.
Futhermore, we note that the derivative of the first term in (\ref{ul_opt})
with respect to $ \lambda_i $ is always non-positive for any 
value of $ \lambda_i $ . Therefore, a nesessary condition that a maximum 
of function $ F(\lambda) $ is attained at a finite value of $ \lambda_i $
is that the derivative of the second term in (\ref{ul_opt}) should be 
non-negative. This imposes a restriction on the range of possible values
of $ \lambda_i $:
\bea
\label{lambda_restr}
\lambda_i & \leq C_{i}/ \sigma_{0}^2   &  {\mbox (DF)} 
\nonumber \\
\lambda_i & \leq - \frac{ 2 ( 1 - \alpha) }{ C_i}  &  {\mbox (CP)} 
\eea
We thus have reduced the problem of minimization of the effective Lagrangian
(\ref{eff_Lagrangian}) to a much simpler finite dimensional optimization 
problem (\ref{NR_problem_ext})-(\ref{solut}) for the Lagrange multipliers 
$ \{ \lambda_i \} $, that may be solved by a multidimensional Newton-Raphson 
method. The computational cost of the method is therefore the same as that of 
the canonical MCE approach.

\section{Adding Intermediate Time Levels}

The last stage of our construction is a calculation of conditional 
transition densities involving additional time levels between the shortest and the 
longest 
maturities of European options in the calibrating set. In this section we
will show how this problem can be consistently addressed using the MCE 
approach. 

In what follows, we will concentrate on the case when a single 
time level $ T $ 
is added between two adjacent maturity dates ($  T_{k} < T < T_{k+1} $) 
in the calibrating set.  
The distributions 
$ p(x), \, p(z|x), \, p(z) $ of asset's prices $ X $ and $ Z $
at times $ T_{k} $ and $ T_{k+1} $, respectively, are 
supposed to be calculated from the 
data as described in the previous sections. We will also need prior densities 
$ q(y|x), q(z|y), q(z|x) $ which may be estimated from the 
$ p $ probabilities found previously by shifting 
time arguments as described above.
We want to calculate the 
distribution $ 
p(y|x) \equiv p(y,T| x,T_k) $, $ p(z|y) \equiv p(z,T_{k+1}| y,T) $. 
Having solved this problem, we could continue with the same method 
recursively and fill the model for all additional time levels of interest. 

One simple-minded approach to the problem would be to say that as soon as 
we know that $ p(y|x) \rightarrow \delta(y-x) $ as $ T \rightarrow T_{k} $
and $ p(y|x) \rightarrow p(z|x) $ as $ T \rightarrow T_{k+1} $, we could 
use linear interpolation to estimate transition densities 
for intermediate times:
\bea
\label{interpol}
\tilde{p}(y|x) &=& (1 - \rho) \, N_{y} (x, \rho) + 
\rho \, p(z|x)|_{z = y} \nonumber \\
\tilde{p}(z|y) &=& (1 - \rho) \, p(z|x)|_{x = y} + \rho \, N_{z} (y, 1- \rho)  
\eea
where $ \rho = (T- T_{k})/(T_{k+1} - T_k) $ and $ N_{x}(m, \sigma^2) $ 
stands for the 
normal distribution in $ x $ with mean $ m $ and variance $  \sigma^2 $. 
A problem with this approach is that such 
approximate transition densities would generally violate 
the Markov 
property
\beq
\label{markov}
p(x,y,z) = p(x) p(y|x) p(z|y)
\eeq
and therefore would not satisfy the Chapman-Kolmogorov
equation
\beq
\label{CK}
\int dy \, p(y|x) p(z|y) = p(z|x) 
\eeq 
and 
the consistency equation 
\beq
\label{consistency}
\int dy \, p(y) p(z|y) = p(z) 
\eeq 
Instead of trying to interpolate, we want to approach 
the problem using the Minimum Cross Entropy approach. 
This amounts
to insisting that the 
asset price distribution should possess a minimum possible relative 
entropy with a reference density for arbitrary times, not only for 
maturity dates of options in the calibrating set. Within this framework,  
equation (\ref{CK})  
can be imposed as a constraint on the MCE solution. The joint KL entropy 
that should be minimized is 
\beq
\label{Joint_entr}
D [p(x,y,z) || q(x,y,z) ] = D[p(x) || q(x) ] + D [p(y|x) || q(y|x) ] + 
D [ p(z|y) || q(z|y) ]
\eeq
where we used the chain rule and the Markov property.
To find the MCE distributions $ p(y|x) $ and $  p(z|y) $, the entropy 
(\ref{Joint_entr}) should be minimized with the Chapman-Kolmogorov
equation 
(\ref{CK}) enforced as a constraint, 
and distributions $ p(x) $  and $ p(z|x) $ fixed. Writing 
the Lagrange multiplier as $ p(x) \lambda(x,z) $ (which we may do as $ p(x) $ is 
fixed), we arrive at the Lagrangian
\bea
\label{triple_lag}
&& L = D [p(x,y,z) || q(x,y,z) ] - \int dx dz \, p(x) \lambda(x,z) \left( 
\int dy \, p(y|x) p(z|y) - p(z|x) \right) \nonumber \\
&&  = \int dx dy dz \, p(x,y,z) \log \frac{ p(x,y,z)}{q(x,y,z)} - 
\int dx dz \, \lambda(x,z) \left( \int dy \, p(x,y,z) - p(x,z) \right)
\eea
Here we used the Markov property (\ref{markov}) to proceed to the second 
expression. 
Minimizing this Lagrangian with respect to $  p(x,y,z) $, we obtain
\beq
\label{triple_res}
p(x,y,z) = \frac{q(x,y,z)}{q(x,z)} \, p(x,z)
\eeq
Note that it is 
crucial for our method that the density $ q $ is non-uniform (at least) 
in the $ y $ direction, 
as otherwise equation (\ref{triple_res}) would indicate that $ p(x,y,z) $
is uniform in $ y $. This is reasonable because in such a 
case there would be no either
prior information (via a reference density), or constraints on 
a distribution of $ Y $. The present method thus adjusts 
the prior distributions
$ q(y|x) $ and $  q(z|y) $  by enforcing the correct joint distribution of the 
end-point values $ p(x,z) $.  

Using the Markov property of the reference distribution, we find the 
resulting conditional densities
\beq
\label{res_y}
p(y|x) = q(y|x) \, \int dz \, \frac{q(z|y)}{q(z|x)} p(x,z) \, 
  \left[ \int dy dz \, \frac{q(y|x) q(z|y)}{q(z|x)} p(x,z) \right]^{-1}
\eeq
\beq
\label{res_z}
p(z|y) = q(z|y) \, \int dx \, \frac{q(y|x)}{q(z|x)} p(x,z) \left[ 
  \int dx dz \, \frac{q(y|x) q(z|y)}{q(z|x)} p(x,z) \right]^{-1}
\eeq
Equations (\ref{res_y}), (\ref{res_z}) solve the problem of finding 
the minimum cross entropy conditional densities $ p(y,T|x,T_{k}) $ and $ 
p(z, T_{k+1}| y,T; x, T_{k}) $, 
given reference densities $ q(y|x),  q(z|y), q(z|x) $ and fixed 
transition density $ p(x,z) $. 
If necessary, the same procedure can now be 
continued to get a more detailed partition of the interval $ [T_{k}, T_{k+1} ]
$ using the same approach.

\section{Numerical examples}

To illustrate the performance of our method, we generate an artificial
``data'' with a Monte-Carlo simulation. We will only consider a calculation
at one time step of our scheme, and moreover use only the DF prior.
A more detailed numerical analysis of our algorithm will be given elsewhere. 
We take $ x_0 = 3.0 $ to be today's ($t =0 $) asset price. We consider $ N = 
5 $ call options with strikes 
\[ 
K = [2.0, 2.5, 3.0, 3.5, 4.0 ] 
\]  
Options mature at $ T = 1 $ year. Risk-free interest rate is $ r = 0.1 $. 
We assume that the ``true'', noiseless prices are given by the Black-Scholes 
model with 
volatility $ \sigma_{BS} = 0.25 $. This yields the following prices 
\[
C =  [1.19, 0.78, 0.45, 0.23, 0.11 ] 
\] 
We model the market noisy prices $ 
\left\{ \hat{C}_{i} \right\} $  as follows: 
\beq
\hat{C}_{i} = C_i \left[ 1 + \sigma_{ns} \, N(0,1) \right]
\eeq
where $ \sigma_{ns} $ measures the noise in the data, and $  N(0,1) $ is a 
random number drawn from a normal distribution with zero mean and unit 
variance. In our simulation we have chosen $ \sigma_{ns} = 0.15 $  
(i.e. 15 \% noise in prices). We have considered the following set of 
``market prices'' obtained with such simulation:
\[
\hat{C} = [1.07, 0.68, 0.53, 0.22, 0.10]
\]
Using this data as an input, we want to calculate with our method the 
transition density $ p(y,T|x,t) $ for $ t = 0.5 $. For the cumulative density
$ p(x) $ at time $ t = 0.5 $, we will assume the lognormal distribution with
the same ``true'' value $ \sigma_{BS} = 0.25 $, while for the prior transition
density $ q(y|x) $ we take the lognormal distribution with $ \sigma_{BS} = 
0.35 $. For the effective parameter $ \sigma \sqrt{h} $ controlling the 
relative weights of the entropy and likelihood terms, we have 
found that the interval $ \sigma \sqrt{h} = [0.45, 3.0] $ provide a 
``stability plateau'' between overfitting (for lower values of $
\sigma \sqrt{h} $) and oversmoothing (for higher values of $ \sigma \sqrt{h} $).
 Results for $ \sigma \sqrt{h} = .45$ and  
$ \sigma \sqrt{h} = 3.0 $ are shown on Figs.1 to 2 and Figs. 3 to 4, 
respectively. Figs. 5 and 6 illustrate the appearance of a fake binodal 
structure in the implied densities at a low value $ \sigma \sqrt{h} = 0.1 $.
For lower values of  $ \sigma \sqrt{h} $, the algorithm blows up, in agreement
with results by Buchen and Kelly \cite{BK}. This shows directly that juggling 
of an implied distribution in the canonical MCE method is due to overfitting.

\begin{figure}
\resizebox{12cm}{!}
  {\includegraphics{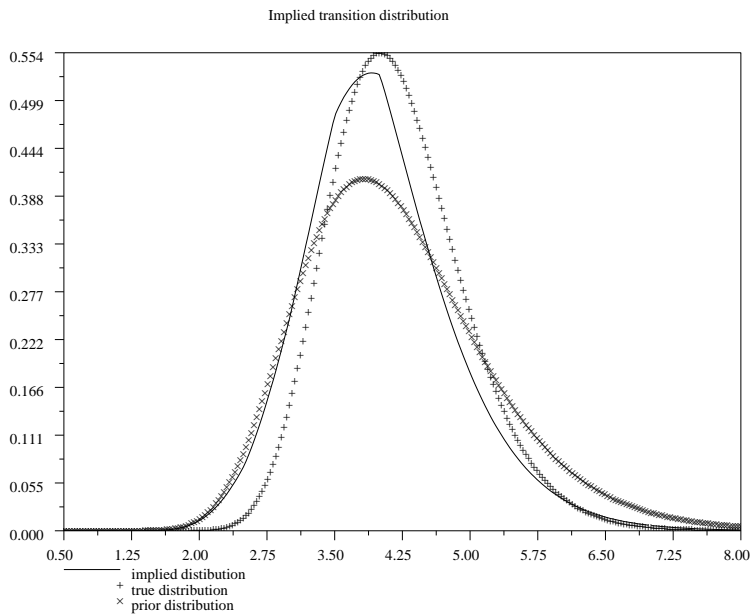}}
\caption{ Implied trandition density calculated with the 
DF prior. $ \sigma \sqrt{h} = 0.45 , x = 4.0 $.
} 
\label{Fig.1}
\end{figure}

\begin{figure}
\resizebox{12cm}{!}
  {\includegraphics{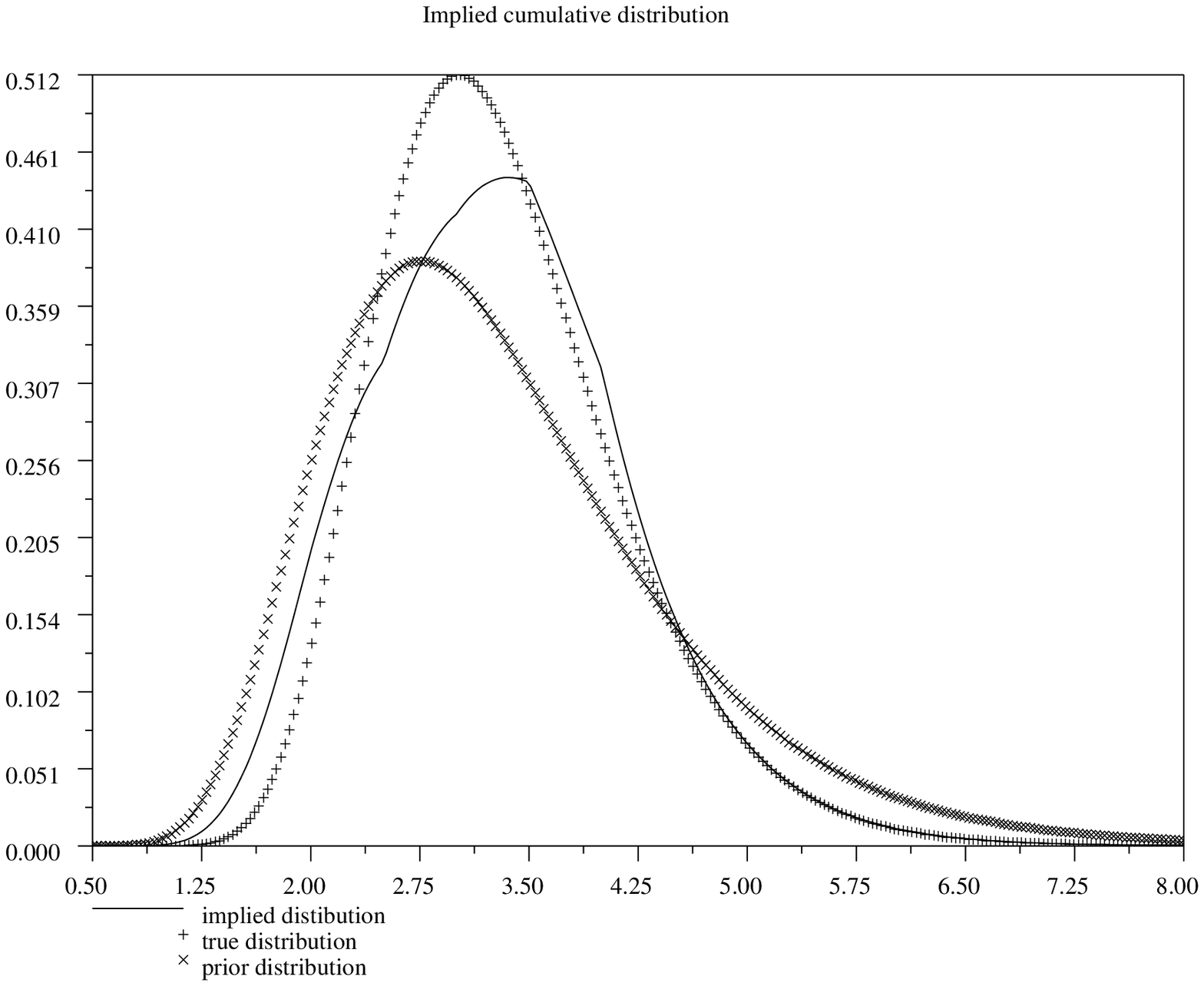}}
\caption{Implied cumulative density calculated with the 
DF prior. $ \sigma \sqrt{h} = 0.45 $. 
} 
\label{Fig.2}
\end{figure}

\begin{figure}
\resizebox{12cm}{!}
  {\includegraphics{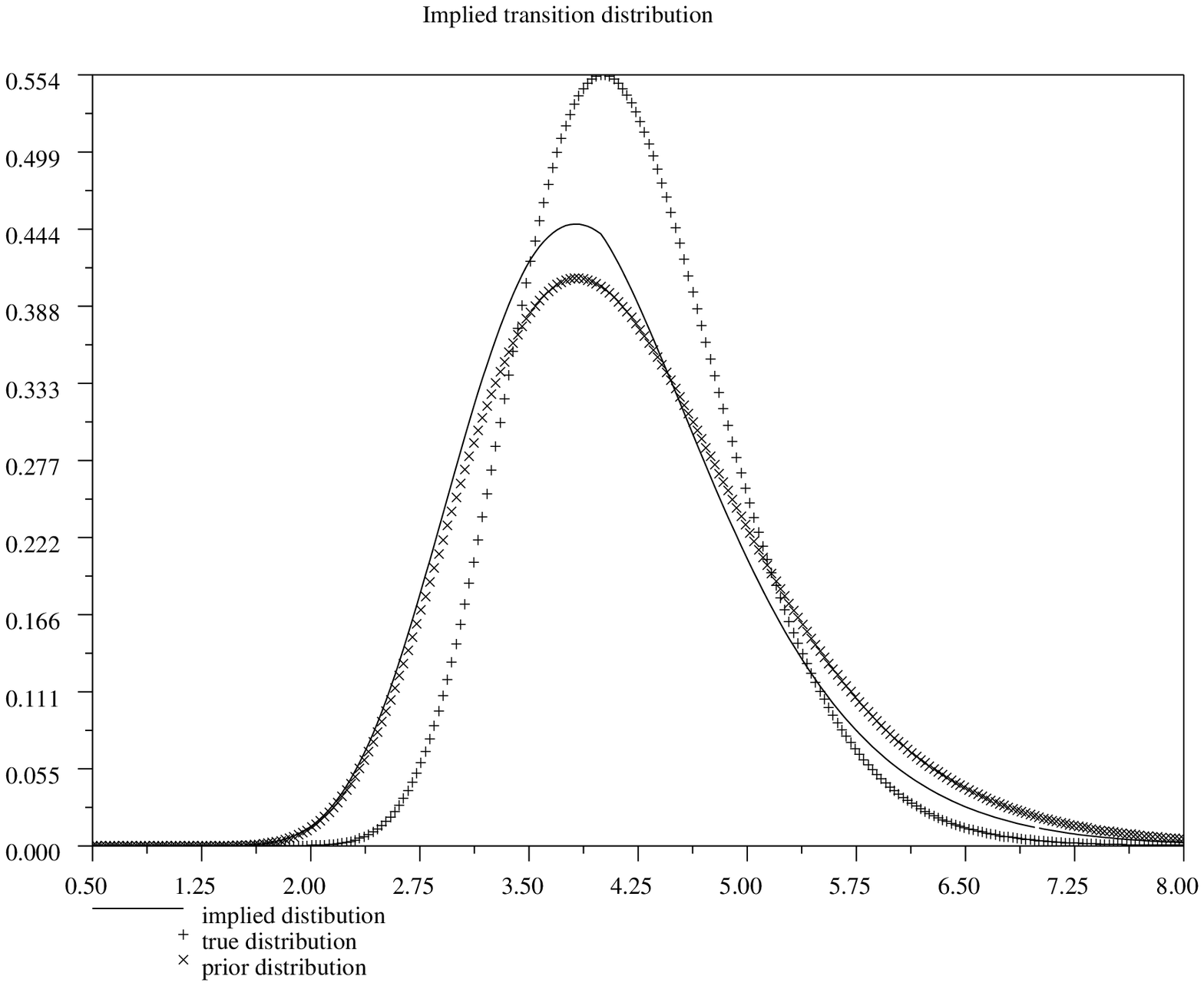}}
\caption{ Implied trandition density calculated with the 
DF prior. $ \sigma \sqrt{h} = 3.0 , x = 4.0 $.
} 
\label{Fig.3}
\end{figure}

\begin{figure}
\resizebox{12cm}{!}
  {\includegraphics{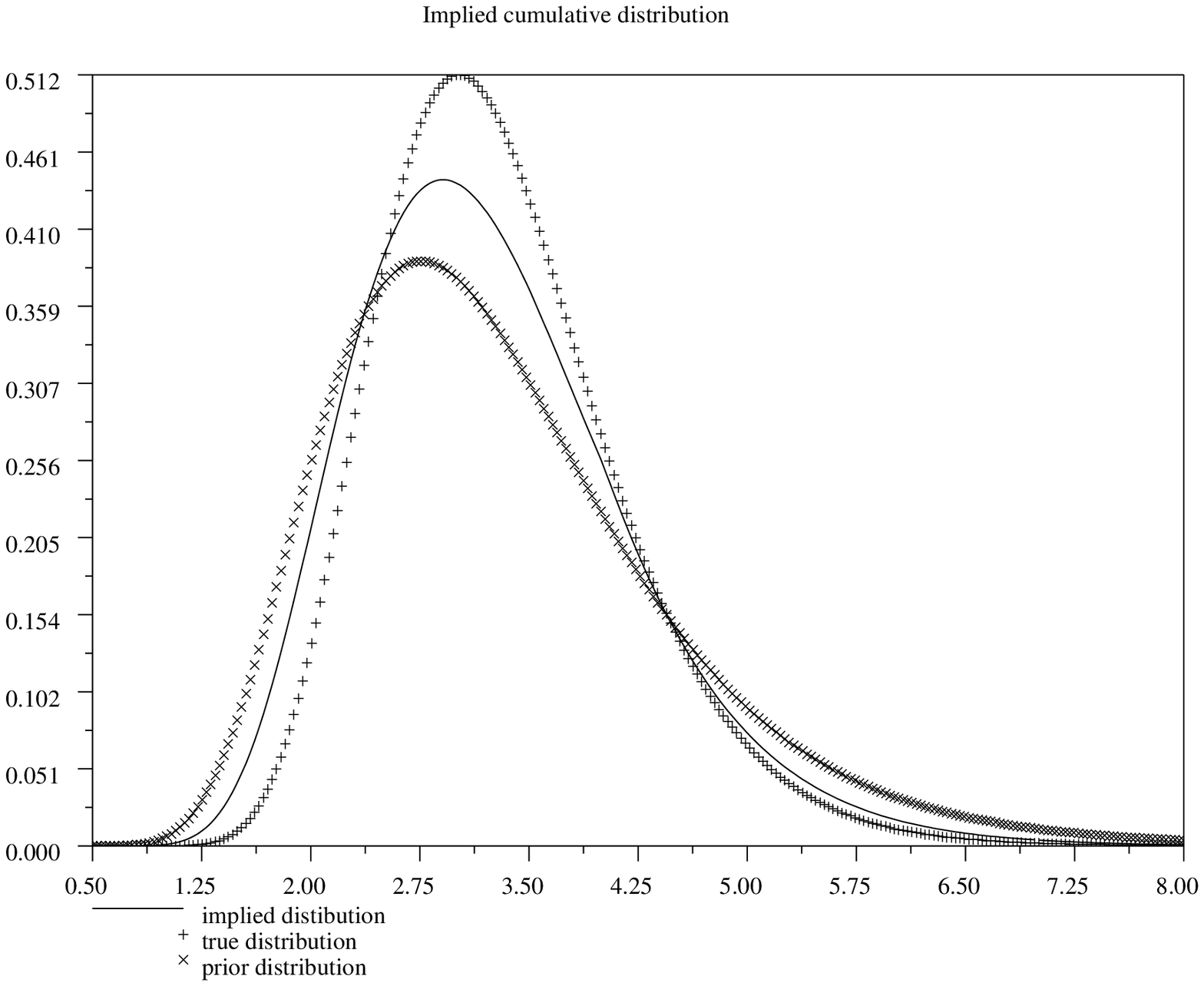}}
\caption{Implied cumulative density calculated with the 
DF prior. $ \sigma \sqrt{h} = 3.0 $. 
} 
\label{Fig.4}
\end{figure}

\begin{figure}
\resizebox{12cm}{!}
  {\includegraphics{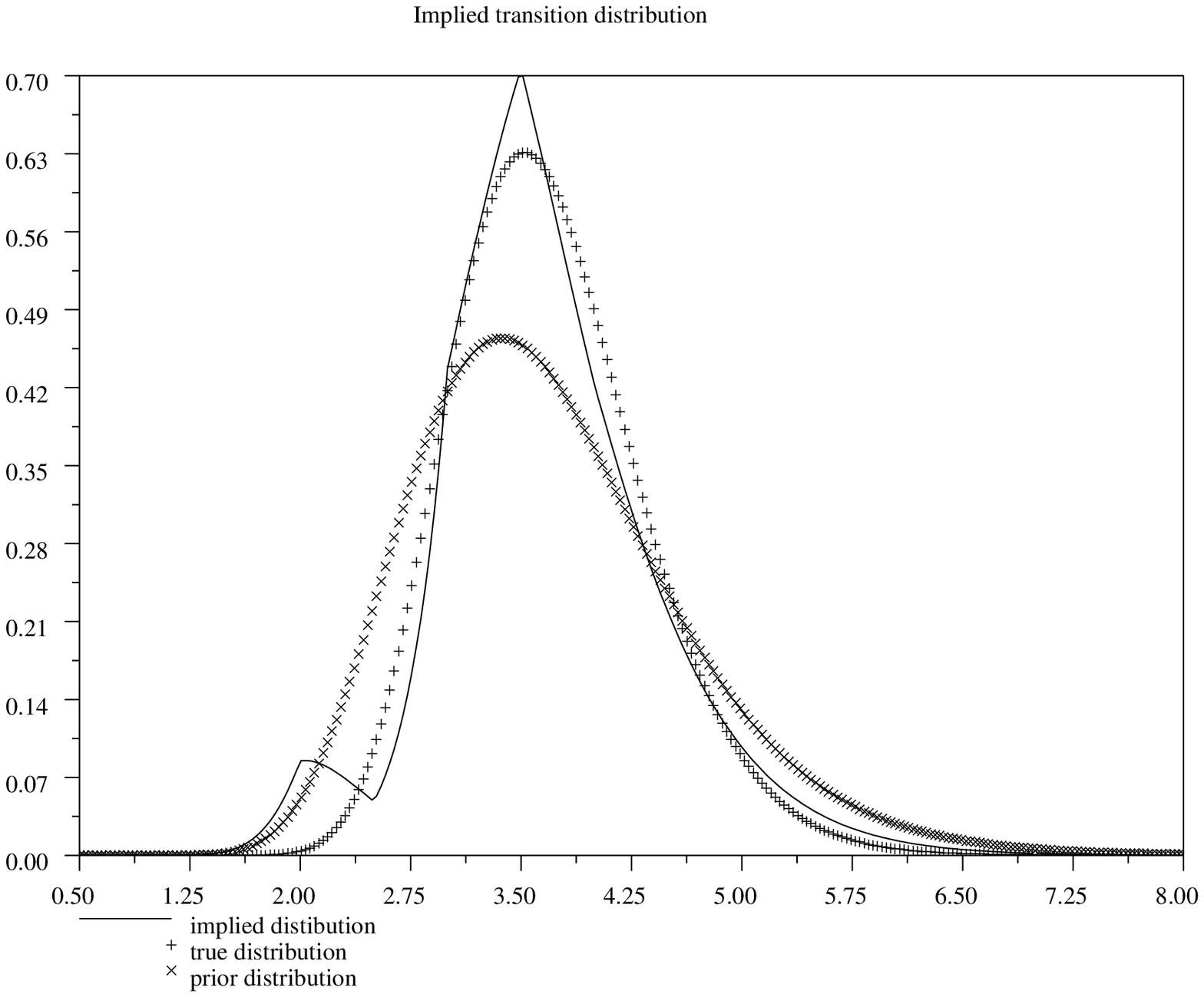}}
\caption{ Implied trandition density calculated with the 
DF prior. $ \sigma \sqrt{h} = 0.1 , x = 3.5 $.
} 
\label{Fig.5}
\end{figure}

\begin{figure}
\resizebox{12cm}{!}
  {\includegraphics{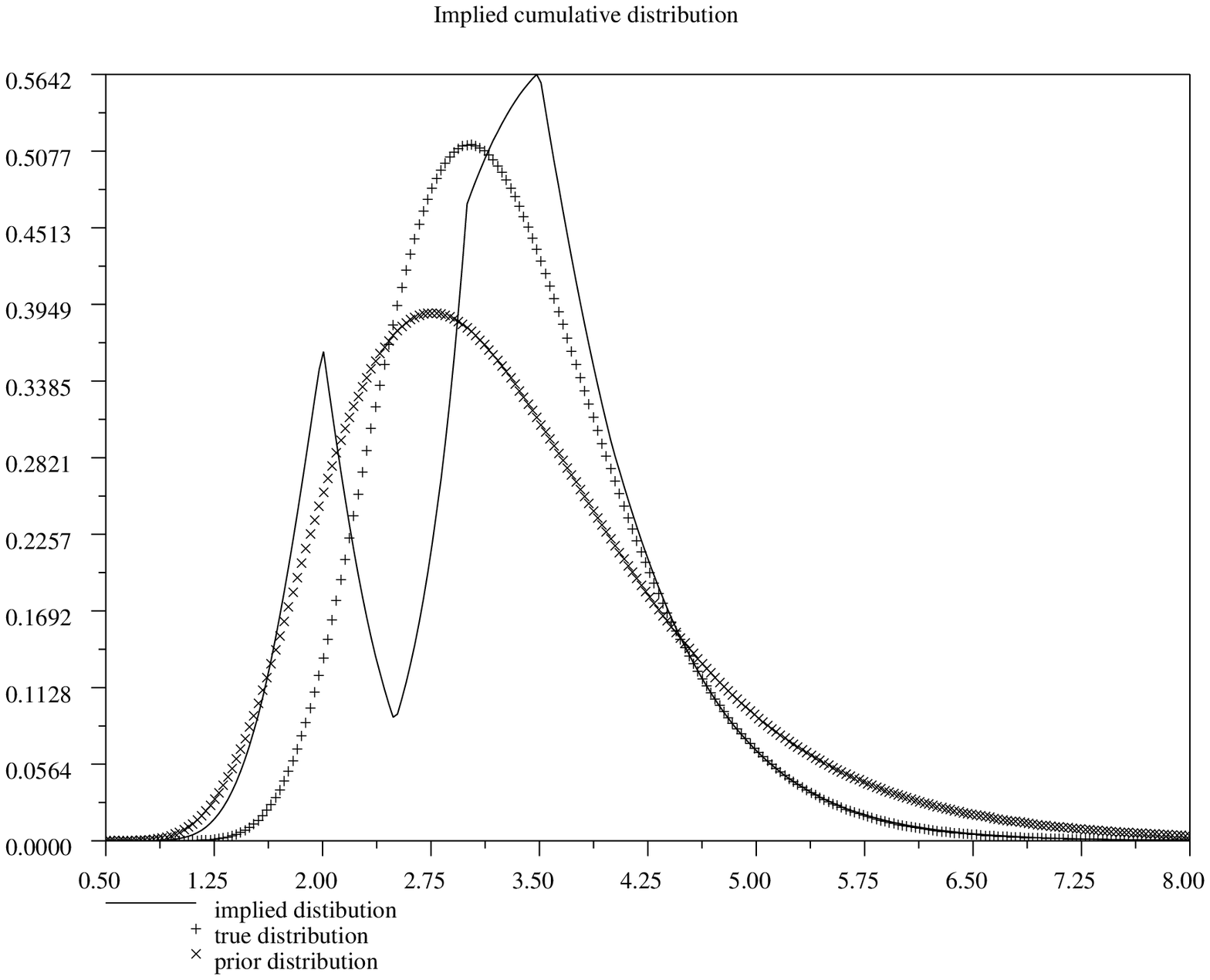}}
\caption{Implied cumulative density calculated with the 
DF prior. $ \sigma \sqrt{h} = 0.1 $. 
} 
\label{Fig.6}
\end{figure}


\end{document}